\documentclass[letterpaper]{article} 
\usepackage{aaai25}
\nocopyright
\usepackage{amsmath}

\usepackage[T1]{fontenc}
\usepackage{times}
\usepackage{booktabs}
\usepackage{multirow}
\usepackage{helvet}
\usepackage{courier}
\usepackage[hyphens]{url}
\usepackage{subfigure}
\usepackage[table,xcdraw]{xcolor}
\usepackage{graphicx}
\graphicspath{{figures/}}
\usepackage{natbib}
\usepackage{caption} 
\ifdefined
\fi

\title{Decision-Centered Evaluation of Machine Learning Poverty Maps Using Mobile Phone and Satellite Data}

\author{
    Chanuka Algama\textsuperscript{1},
    Merl Chandana\textsuperscript{1},
    Viren Dias\textsuperscript{1,2},
    Kasun Amarasinghe\textsuperscript{1,3}
}
\affiliations{
    \textsuperscript{1}LIRNEasia,
    \textsuperscript{2}Calcey,
    \textsuperscript{3}Carnegie Mellon University\\
    \{chanuka, merl\}@lirneasia.net\\
    viren@calcey.com\\
    kamarasi@andrew.cmu.edu
}

\begin{document}

\maketitle

\begin{abstract}
Identifying the poorest communities is essential for poverty alleviation, but household surveys and censuses are costly and infrequent. Machine learning offers alternative poverty estimates from mobile phone and satellite data, yet average prediction accuracy alone does not show whether maps support targeting under limited budgets. We apply a decision-centered evaluation to 13,985 Grama Niladhari divisions in Sri Lanka, combining call detail records (CDRs), remote sensing (RS), and CNN-derived Landsat 8 embeddings. We assess recovery of the poorest administrative units, compare random and spatially grouped validation, and examine errors in socioeconomically atypical communities. Against a census-derived asset index (PC1), Random Forest achieves Recall@25\% of 0.830, compared with 0.450 for nighttime lights. Random splitting raises recall by 4.1 percentage points relative to Divisional Secretariat Division (DSD)-grouped holdouts. The combined model recovers 86\% of the 25 poorest DSDs by PC1, versus 69\% for RS-only and 67\% for CDR-only models. Spatially isolated divisions have 10.8\% higher prediction error. Stronger isolation--error association in RS-only models is consistent with spatial smoothing, although its cause remains unconfirmed. These findings support evaluating poverty maps by targeting performance and geographic transfer, while recognising that agreement with an asset index does not establish consumption-poverty accuracy.

\end{abstract}

\begin{figure}[h!]
\centering
\subfigure[PC1 reference]{\includegraphics[height=6.7cm]{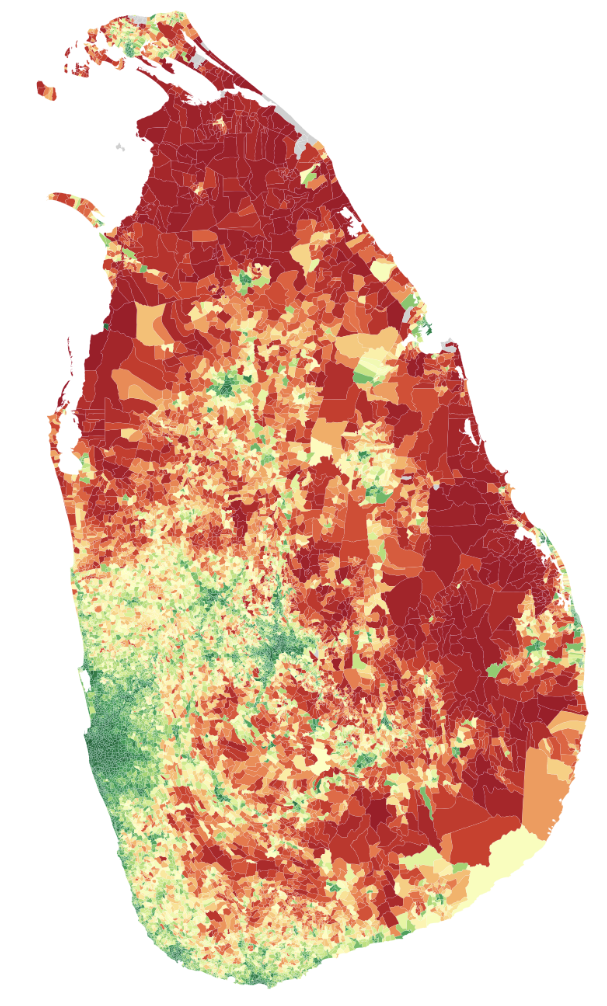}}
\subfigure[ML estimation]{\includegraphics[height=6.7cm]{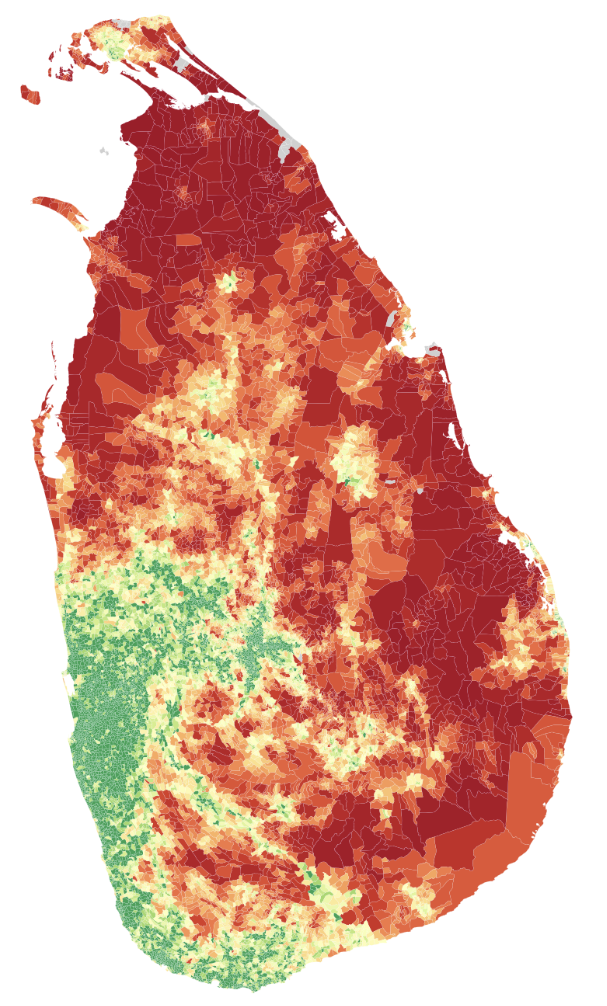}}
\caption{Poverty percentile ranks across 13,985 GN divisions, Sri Lanka's smallest administrative units: (a) census-derived PC1 reference (2012); (b) Random Forest predictions from mobile phone and satellite features. Both panels use the same scale: red denotes greater deprivation and green less deprivation. The north, northeast, and interior highlands appear poorer, while the Western Province corridor appears wealthier (Spearman $\rho=0.8532$ between maps).}

\label{map}
\end{figure}

\section{Introduction}

Poverty continues to impose severe social and economic costs across the globe, contributing to child mortality, reduced access to education, social instability, and conflict \cite{cruz2015ending}. Despite sustained international efforts, poverty still remains as a global challenge. The World Bank's 2024 assessment estimated that 8.5\% of the global population lived below the then-used extreme-poverty line of US\$2.15 per day, with progress slowed by COVID-19 and other shocks \cite{worldbank2024poverty}.

Effective poverty alleviation requires reliable spatial information on where the most vulnerable populations live. Spatially explicit poverty data play a critical role in guiding development interventions, in allocating limited public resources, and monitoring the effectiveness of social protection programs \cite{akinyemi2007spatial}. Governments and development agencies frequently rely on such information when designing targeted initiatives such as infrastructure investments, welfare programs, or emergency aid following economic shocks or natural disasters.

However, obtaining timely and granular poverty estimates remains a major challenge. Decennial censuses provide detailed socioeconomic data, but they are costly, infrequent, and often outdated by the time they become available. Household Income and Expenditure Surveys (HIES), another commonly used instrument in developing countries, provide important insights into consumption and welfare but are similarly resource-intensive and typically conducted only every few years. These traditional measurement approaches therefore struggle to capture the dynamic and geographically heterogeneous nature of poverty, particularly in low and middle income countries such as Sri Lanka. Different welfare measures capture different dimensions of deprivation, so a model's target must be interpreted in relation to the intended programme.

Recent advances in machine learning and the increasing availability of large-scale geospatial data have created new opportunities for poverty mapping. Remote sensing (RS) indicators, satellite imagery, and mobile phone call detail records (CDRs) have each been shown to correlate with socioeconomic conditions at fine spatial resolution. RS data capture environmental and geographic characteristics such as vegetation, infrastructure, settlement density, climate, and nighttime light intensity \cite{engstrom2017poverty, mitterling2021compiling}. Deep learning models applied to satellite imagery have demonstrated strong performance in predicting regional wealth indices \cite{jean2016combining, steele2017mapping}. CDR data provide complementary behavioural information through mobility patterns, communication activity, recharge behaviour, and social network structure \cite{blumenstock2015predicting}. More recent work has combined multiple modalities to generate large-scale poverty estimates at increasingly high spatial resolution \cite{chi2022microestimates, espin2023interpreting, agyemang2023high}.

For geographic targeting, prediction quality must be assessed against the allocation decision. Average-error metrics such as RMSE and $R^2$ remain useful, but do not directly measure inclusion and exclusion at a specified budget. Targeting research has long studied these questions; our focus is their joint application with spatial validation to multimodal ML maps.

Spatial dependence also changes what a validation result means. Nearby training and test units can share socioeconomic conditions, so random splits may overstate performance for deployment to new areas. Spatial blocking is an established response \cite{roberts2017cross,valavi2019blockcv}; administrative grouping is a practical instance, not a guarantee of independence.

\begin{figure*}[t]
    \centering
    \includegraphics[width=15cm]{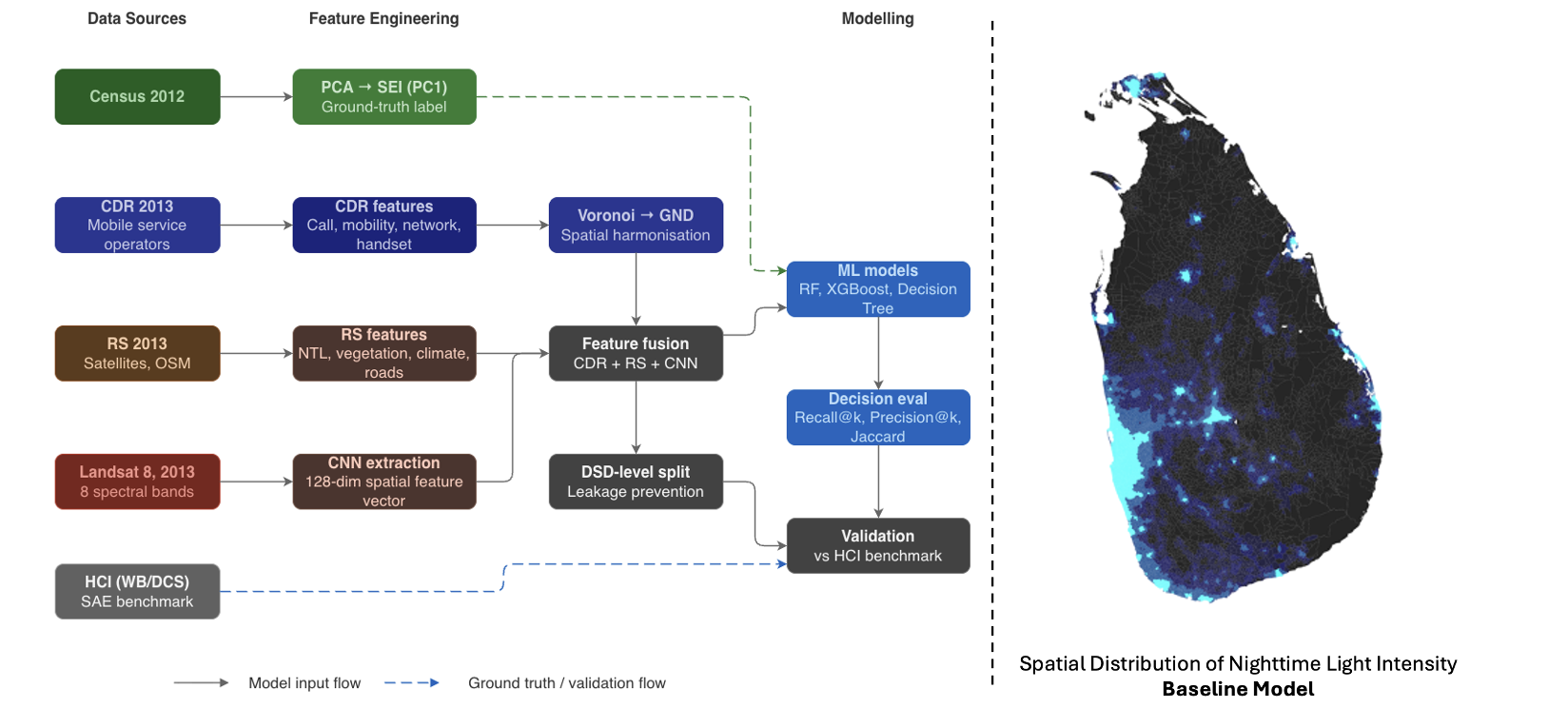}
    \caption{Poverty mapping pipeline and NTL baseline. (Left) The five data sources used in the pipeline: 2012 census data, CDR data, remote sensing indicators, Landsat 8 imagery, and the World Bank / DCS Poverty Headcount Index (HCI). These sources are used across feature engineering, model training, and decision-centered evaluation stages. (Right) VIIRS nighttime light intensity across Sri Lanka used as the NTL ranking baseline, where darker areas indicate lower economic activity and higher poverty likelihood.}
    \label{pipeline}
\end{figure*}

Using Sri Lanka as a case study, we estimate deprivation at the Grama Niladhari (GN) division level, the smallest administrative unit. Our contributions are empirical: (1) applying budget-based ranking evaluation to multimodal poverty maps, while distinguishing asset-proxy agreement from consumption poverty; (2) comparing data modalities and quantifying the sensitivity of reported performance to DSD-grouped and province-held-out validation; and (3) documenting an association between local socioeconomic contrast and prediction error, which we call \textit{spatial isolation bias}. The term describes socioeconomic dissimilarity from neighbours, not geographic remoteness. We do not introduce new ranking metrics, a new spatial cross-validation method, or a new predictive architecture.

\section{Related Work}
\label{sec:related}

\paragraph{Targeting under resource constraints.}
Geographic targeting has long evaluated how limited resources reach disadvantaged populations. \citet{elbers2007poverty} simulate budget-constrained transfers using poverty maps and examine gains from finer geographic disaggregation. \citet{coady2004targeting} review transfer-targeting performance, while \citet{grosh1995proxy} examine proxy means tests and targeting errors. These studies establish the relevance of inclusion and exclusion errors. More recently, \citet{aiken2022machine} evaluate phone-based ML for targeting humanitarian aid. Our set-overlap evaluation belongs to this tradition: it measures recovery of reference-ranked areas, not the welfare gains or household eligibility outcomes assessed by a complete programme evaluation.

\paragraph{Learning poverty proxies from alternative data.}
\citet{blumenstock2015predicting} predict wealth from phone metadata; \citet{jean2016combining} use satellite-image transfer learning; and \citet{steele2017mapping} combine mobile and satellite data. \citet{chi2022microestimates} extend wealth estimation across low- and middle-income countries, and \citet{espin2023interpreting} investigate multimodal wealth inference. Deep architectures are relevant alternatives: \citet{agyemang2023high} ensemble CNNs for fine-resolution rural poverty mapping in Pakistan and validate against survey data. These studies motivate our inputs and comparisons, but their different targets, spatial supports, and validation settings prevent a direct ranking of published accuracies. Our tree models provide an evaluation case study; their performance does not establish superiority over advanced deep models.

\paragraph{Spatial validation.}
\citet{roberts2017cross} explain why validation must reflect dependence and the intended prediction task. \citet{valavi2019blockcv} implement spatial and environmental blocking strategies. We apply these principles using DSD boundaries and province holdouts. DSD grouping prevents within-DSD overlap between training and validation, but dependence across adjacent DSDs can remain. Our empirical contribution is measuring the resulting performance differences in this poverty-mapping setting.

\section{Data Sources and Pipeline Overview}

Our pipeline brings together five data sources: (1) census data to construct an asset-based reference index, (2) CDR data, (3) RS indicators, and (4) raw satellite imagery as the primary model inputs available in the absence of census data, and (5) an externally constructed consumption-poverty benchmark for external validation. 
Figure~\ref{pipeline} illustrates how these sources
interact across the pipeline. The remainder of this section describes each source,
its role, and the rationale for its inclusion.

\begin{table}[t]
\centering
\begin{tabular}{l|l}
\toprule
    Metric Type & Feature \\
    \midrule
    Phone usage & Call count \\
    & Average call duration  \\
    & Geometry \\
    & Nighttime call count \\
    & Incoming call count \\
    & Avg nighttime call duration \\
    & Avg incoming call duration  \\
    & Avg outgoing call duration \\
    Location/Mobility & Radius of gyration \\
    & Home location \\
    Social Network & Spatial entropy \\ 
    & Avg call count per contact \\
    Handset type &  Smart/feature/basic phone \\
    \bottomrule
\end{tabular}
\caption{Features derived from CDR data}
\label{cdrfeatures}
\end{table}

\subsection{Census-Based Socioeconomic Index}
We construct an asset-based socioeconomic index from the 2012 census using principal component analysis (PCA), following established approaches to measuring relative welfare \cite{filmer2001estimating,vyas2006constructing}. Starting with 109 household and demographic variables, including housing quality, utilities, sanitation, and assets, we removed near-constant and redundant features before applying PCA. We retained the first component (PC1) as the GND-level reference target, with lower values indicating greater deprivation. PC1 measures relative asset deprivation rather than consumption poverty; Section~\ref{sec:pc1_validation} compares its DSD aggregates with a consumption-based benchmark.

\subsection{Call Detail Records}
We use 2013 CDRs from two leading Sri Lankan mobile operators to derive indicators of call activity, mobility, social networks, and handset type (Table~\ref{cdrfeatures}). These behavioural measures have been associated with wealth in earlier studies \cite{blumenstock2015predicting,steele2017mapping}.

Records were aggregated at cell-tower level. We approximated tower coverage using Voronoi polygons and assigned each GND the mean, sum, or mode of values from overlapping polygons, depending on the feature. The effective spatial resolution therefore depends on tower density and is generally coarser in rural areas.

\subsection{Remote Sensing and Satellite Imagery}

\begin{table}[t]
\centering
\resizebox{\columnwidth}{!}{
\begin{tabular}{l|l}
\toprule
    Information type & Feature \\
    \midrule
    Accessibility & Accessibility to populated places with more \\
    & than 50k people \\
    Population & Population count \\
    & Population density \\
    Climate & Mean aridity index \\
    & Mean annual precipitation \\
    & Average annual evapotranspiration \\
    & Mean annual temperature \\
    Night-time lights & VIIRS satellite night-time lights intensity \\
    Elevation & Elevation in meters \\
    Vegetation & Vegetation Index \\
    Distance & Distance to roads \\
    & Distance to waterways \\
    Urban/rural & MODIS satellite-based global urban extent \\
    Protected area & Protected areas \\
    Land cover & European Space Agency land cover maps \\
    Demographic & Pregnancies \\
    & Births \\
    Ethnicity & Georeferenced ethnic groups \\
    \bottomrule
\end{tabular}}
\caption{Features derived from GIS \& Remote Sensing Data for the year 2013}
\label{rsfeatures}
\end{table}

We assembled 2013 indicators from open geospatial repositories, satellite products, and OpenStreetMap-derived datasets, one year after the census. Table~\ref{rsfeatures} lists the features, including nighttime lights, vegetation, accessibility, climate, and population. These capture physical correlates of economic activity \cite{henderson2012measuring,engstrom2017poverty,mitterling2021compiling} and provide coverage beyond mobile subscribers, although their spatial resolution varies.

We also extracted image features from radiometrically calibrated and atmospherically corrected Landsat 8 scenes for 2013. Scenes were clipped to GND boundaries to form eight-channel arrays. The CNN had three convolutional blocks with 32, 64, and 128 filters, $3\times3$ kernels, ReLU activations, and $2\times2$ max pooling. A flattening layer was followed by a 128-unit dense layer with dropout and a two-unit output layer. The penultimate layer supplied each GND's 128-dimensional embedding.

The preprocessing records still need to establish the exact band list and treatment of variable image sizes. The CNN target, loss, and fold-specific training procedure also remain undocumented. In particular, we cannot establish whether held-out DSD labels were excluded when learning the embeddings. Downstream spatial splits alone therefore cannot rule out representation-stage leakage; this requires fold-specific supervised training or an independently pretrained, frozen encoder.

\subsection{External Validation Benchmark}
We compare PC1 with the DCS/World Bank poverty headcount index (HCI), which combines the 2012 census and 2012/13 Household Income and Expenditure Survey through small area estimation at DSD level \cite{DCS2015PovertyMap}. HCI measures consumption poverty but shares census inputs with PC1 and carries estimation uncertainty. The comparison assesses PC1--HCI agreement; direct model--HCI targeting accuracy is not reported.

\subsection{Assessing PC1 Against Consumption Poverty}
\label{sec:pc1_validation}

\begin{figure*}[h]
    \centering
    \includegraphics[width=\textwidth]{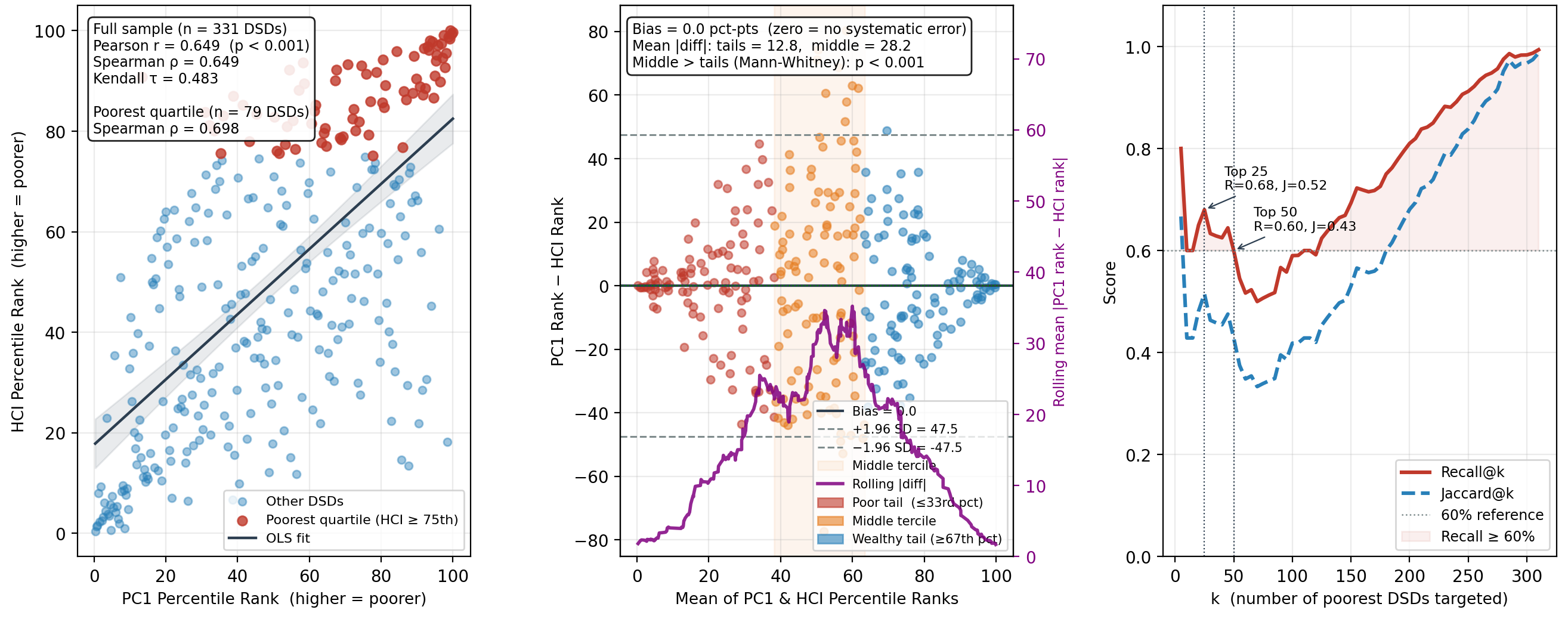}
    \caption{Validation of PC1 as a poverty proxy against the World Bank / DCS
    Poverty Headcount Index (HCI) across 331 DSDs. (A) DSD-level rank correlation
    between population-weighted PC1 percentile ranks and HCI percentile ranks
    (Pearson $r=0.649$); the highlighted HCI-poorest quartile has Spearman $\rho=0.698$. (B) Differences between percentile ranks. The zero mean follows from using ranks of the same units and is not evidence of absence of systematic error, despite the original panel annotation. (C) Recovery of HCI-poor DSDs by PC1: Recall@25 DSDs $=0.68$ and Jaccard $=0.52$. These quantify proxy agreement, not model performance.}

    \label{pcaVShci}
\end{figure*}

To assess whether PC1 provides an acceptable ordinal representation of poverty for
our ranking-based framework, we aggregated GND-level PC1 scores to the DSD level
using population-weighted means and compared them against the HCI. Across all 331
DSDs, PC1 and HCI percentile ranks had Pearson and Spearman correlations of 0.649. The originally reported Pearson 95\% interval [0.579, 0.708] does not establish spatially adjusted uncertainty.

The mean signed difference between percentile ranks is 0.00, but this follows mechanically when both rankings cover the same units with the same rank convention. It does not demonstrate unbiased poverty measurement. The mean absolute rank difference was 12.8 percentile points in the combined tails and 28.2 in the middle tercile (Figure~\ref{pcaVShci}). This is descriptive evidence of heterogeneous agreement, not validation of every poor area. Asset holdings and current consumption capture different aspects of welfare and may diverge.

At the 25-poorest-DSD threshold, PC1 recovers 68\% of HCI-selected DSDs (17 of 25; Jaccard $17/33\simeq0.52$). Thus nearly one-third of the consumption-poor reference set is missed even before model error is introduced. High model--PC1 recall measures faithful recovery of this proxy; it does not establish equivalent recovery of consumption-poor DSDs or households. Direct model--HCI evaluation and sensitivity to alternative reference indices are needed to assess that downstream uncertainty.

\section{Model Building and Validation}

\subsection{Binary Poverty Label}

We train classifiers using $y_i=\mathbf{1}[\mathrm{PC1}_i<-3]$, which labels 1,925 GNDs (13.8\%) as positive. This empirical asset-index cut-off is not an official poverty line. The fraction of administrative units below it is also not a population poverty headcount, and a DSD-level rank comparison cannot validate a particular GND-level cut-off.

The training cut-off and evaluation budget are separate: classifiers learn this fixed label, while evaluation compares their score rankings with the bottom $k$ units by continuous PC1. Varying $k$ in a recall curve does not test sensitivity to the training cut-off. Such sensitivity requires refitting under alternative PC1 thresholds on the same spatial splits; it has not been established here.

\subsection{Classifiers}

We trained three ensemble and tree-based classifiers: Random Forest (1,000 trees), XGBoost (300 trees, maximum depth 6, learning rate 0.05), and a Decision Tree (maximum depth 5). All classifiers were trained to output predicted probabilities of poverty rather than binary class labels, as the probability score is used directly to
rank GN divisions from most to least likely to be poor. This ranking is the primary model output evaluated in all subsequent analyses.

We evaluated three feature configurations for each classifier: (1) the full multimodal input (CDR + RS + CNN-derived embeddings), (2) RS features only, (3) and CDR features only.
These modality comparisons describe performance under different input choices. They do not separately isolate the incremental contribution of CNN embeddings, which would require a CDR+tabular-RS comparison with and without embeddings on identical splits.

We use tree-based classifiers to study nonlinear combinations of tabular features and image embeddings with a common ranking output. A shallow Decision Tree supplies a simpler model comparison. This scope supports an evaluation study, not a claim of state-of-the-art predictive performance; comparisons with end-to-end CNNs, pretrained encoders, and spatial architectures remain necessary.

\subsection{Spatial Weights for Prediction Error Analysis}

For supplementary inference analyses examining which covariates are most associated with poverty, and to test for spatially structured prediction errors (see Figure~\ref{isolation}), we constructed a spatial weights matrix assigning each GN division a fixed set of $k$ nearest neighbours based on geographic proximity. 
Sri Lanka contains 62 island GN divisions that share no contiguous boundaries with any other unit, making contiguity-based spatial weights infeasible. The $K$-nearest-neighbour (KNN) construction, distinct from the CNN image encoder, ensures that all units receive valid spatial weights regardless of adjacency, which is essential for spatial regression and spatial autocorrelation analysis.

\subsection{Validation Framework}

\subsubsection{Matching Validation to Geographic Transfer}
A random GND split can place highly similar neighbours in training and validation. The resulting score can be optimistic for prediction in new geographic areas, although interpolation near observed units is itself a distinct prediction task. We compare random and DSD-grouped splits in Section~\ref{leak}, following established spatial-validation principles \cite{roberts2017cross,valavi2019blockcv}.

\subsubsection{Repeated DSD-Grouped Holdout}
At each repetition, 80\% of DSDs are assigned to training and 20\% to validation, keeping every GND within a DSD in the same partition. The reported procedure repeats this assignment 1,000 times. It is repeated grouped holdout, rather than a classical bootstrap: sampling DSDs with replacement is not specified. We use this terminology throughout.

Grouping reduces within-DSD dependence across partitions, but adjacent DSDs may remain correlated. Shared tower coverage and learned representations can also cross boundaries. Administrative grouping therefore does not establish complete absence of leakage. Buffered spatial holdouts and checks of residual autocorrelation would further assess separation. Variation over repeated splits measures sensitivity to partition assignment, not uncertainty from new census data, the poverty proxy, or temporal change.

\subsubsection{Province Leave-One-Out Cross-Validation}

To assess geographic generalisability, that is, whether models trained on one part of
Sri Lanka perform acceptably in unseen regions, we conduct a leave-one-province-out
evaluation. Sri Lanka is divided into nine provinces with distinct geographic,
demographic, and economic characteristics. At each fold, one province is withheld
entirely as the test set and the model is trained on the remaining eight provinces.
This evaluation directly answers whether the learned relationships between CDR/RS
features and poverty generalise across the country's varied geographic contexts, from
the densely populated Western Province to the conflict-affected Northern Province and
the rural agricultural provinces of the interior.

\subsubsection{Decision-Centered Evaluation Metrics}

Our primary evaluation framework is decision-centered. Rather than measuring how
closely model scores reproduce reference values everywhere, we measure how
well models support the specific policy task of identifying and targeting the poorest
GN divisions under a finite resource budget.

Let $\hat{S}_k$ contain the $k$ evaluated units with highest predicted poverty scores and $S_k^*$ the $k$ evaluated units with lowest PC1. The same evaluation units and cardinality are used for both sets. Then
\begin{equation}
\begin{aligned}
\operatorname{Recall}@k &= \frac{|\hat{S}_k\cap S_k^*|}{|S_k^*|},\\
\operatorname{Precision}@k &= \frac{|\hat{S}_k\cap S_k^*|}{|\hat{S}_k|},\\
\operatorname{Jaccard}@k &= \frac{|\hat{S}_k\cap S_k^*|}{|\hat{S}_k\cup S_k^*|}.
\end{aligned}
\label{eq:targeting}
\end{equation}
Here $|S_k^*|=|\hat{S}_k|=k$, so Precision@$k$ equals Recall@$k$ exactly, and Jaccard@$k=R_k/(2-R_k)$ within each split, where $R_k$ is recall. These are alternative summaries of the same overlap, not independent evidence. At fixed $k$, $1-R_k$ is both the exclusion fraction from the reference set and the inclusion-error fraction among selected units. A fixed poverty-line reference set of size different from $k$ would yield different precision and recall.

We evaluate fractional budgets of 5\%, 10\%, $\ldots$, 100\% of evaluated units, reporting means and 10th--90th percentile bands over repeated splits. Fixed-count thresholds, such as 25 DSDs or 100 GNDs, are explicitly labelled as counts. Rounding and score-tie handling must be fixed consistently across methods for reproducibility. Random selection has expected recall $k/N$ on $N$ evaluated units, providing a budget-specific reference.

Selecting $k$ units represents a budget only under an equal-cost-per-unit assumption. These unweighted area metrics do not measure the number of poor people reached, intervention cost, poverty reduction, or household eligibility. Those quantities require population weights, costs, and programme-specific outcomes.

\subsubsection{Baseline Models}

As reference points for comparison, we built two simpler ranking approaches that
yield an ordinal mapping of GNDs. The first ranks GN divisions by raw VIIRS nighttime light (NTL) intensity in descending order of darkness, assuming that lower NTL corresponds to greater poverty. The second ranks GN divisions by population density in descending order, assuming denser areas tend to be poorer. Both baselines represent information that is freely and immediately available to policymakers.

\section{Results}

\subsection{Recovery of the PC1 Poor Tail}

\begin{figure}[t!]
    \centering
    \includegraphics[width=\columnwidth]{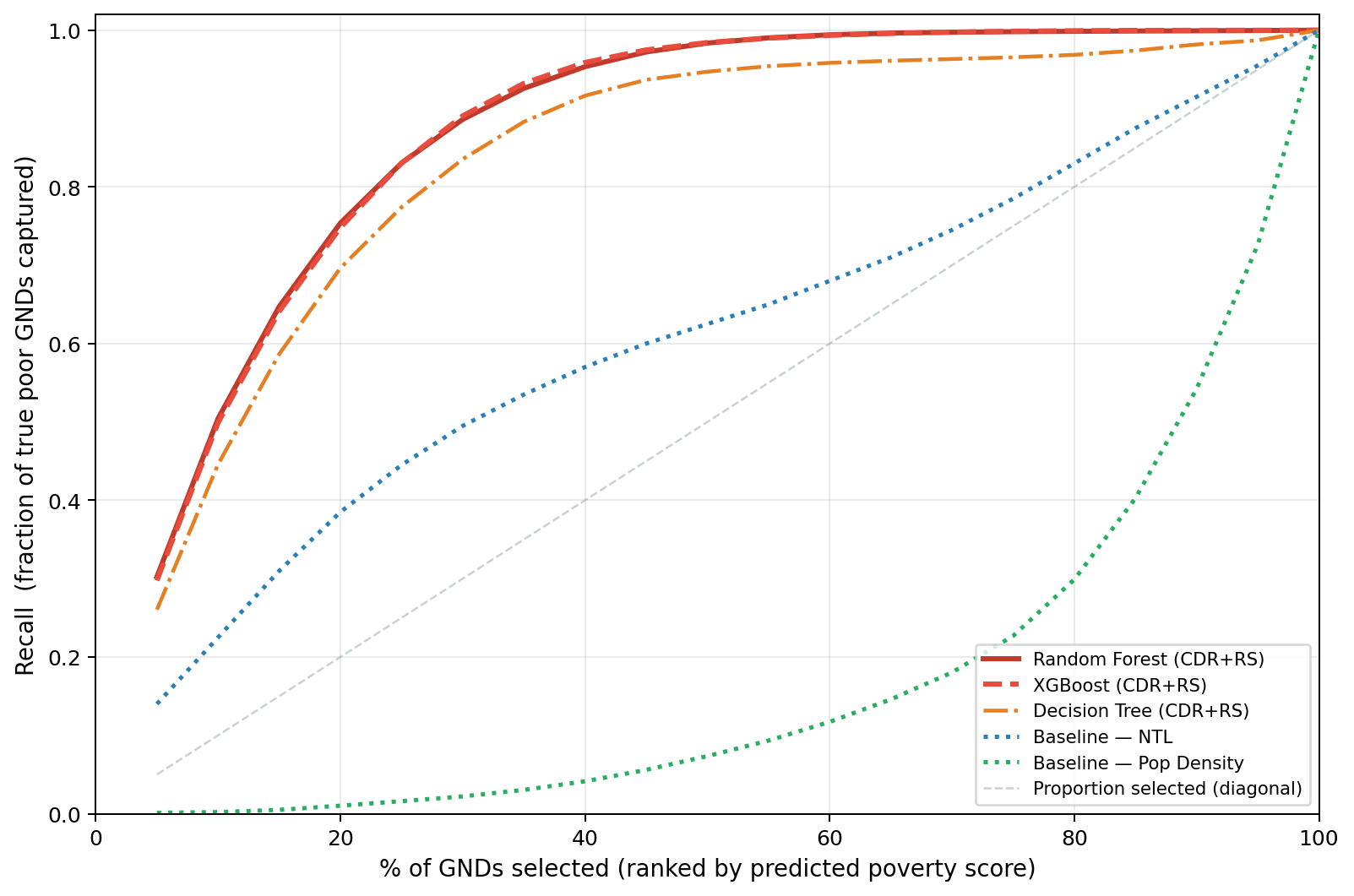}
    \caption{Mean Recall@$k$ curves for Random Forest, XGBoost, and Decision Tree
    (CDR+RS) against NTL and population density baselines, averaged across 1,000
    DSD-grouped holdout repetitions. Random Forest and XGBoost recover 83\% of the PC1 reference set when
    targeting the poorest 25\%, compared to 45\% for NTL and below-random performance
    for population density.}
    \label{recall}
\end{figure}

Figure~\ref{recall} shows mean Recall@$k$ curves across 1,000 DSD-grouped holdout repetitions for all models and both baselines. At the policy-relevant threshold of targeting the poorest 25\% of GN divisions, a budget under which a hypothetical intervention could serve approximately 3,500 GNDs, Random Forest and XGBoost each achieve Recall@25\% $= 0.830$, recovering 83\% of the PC1-bottom-quartile reference set. This is also Precision@25\% under the equal-cardinality definition. The Decision Tree achieves 0.774 at the same threshold.

The NTL baseline reaches Recall@25\% $=0.450$, recovering 45\% of the PC1 reference set compared with 83\% for Random Forest. The net recall advantage is 38 percentage points at the same area budget. Population density ranked in descending order performs below random selection
(Recall@25\% $=0.016$ versus an expected $0.25$ for random selection). This result
shows that the chosen ranking direction is poorly aligned with PC1; it does not rule
out a reversed or learned relationship with population density.
All model recall curves rise steeply in the critical early region of the distribution
and converge toward 1.0 as the targeting budget expands.

\subsection{Random Splits Yield Higher Performance Estimates}
\label{leak}

\begin{figure}[t!]
    \centering
    \includegraphics[width=\columnwidth]{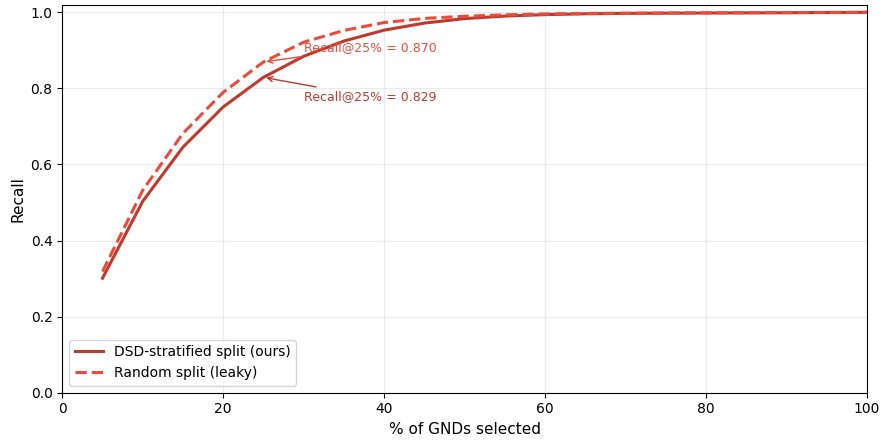}
    \caption{Validation comparison of random GND-level splitting against
    DSD-grouped splitting (Random Forest, 200 iterations). Random splitting
    inflates Recall@25\% by 4.1 percentage points (0.870 vs 0.829), showing sensitivity to
    the geographic separation used in validation.}
    \label{leakage}
\end{figure}

Figure~\ref{leakage} demonstrates the consequence of using random GND-level splits
rather than our DSD-grouped approach. Under random splitting, the same Random
Forest model achieves Recall@25\% $= 0.870$, which is 4.1 percentage points higher
than the 0.829 obtained under DSD-grouped splitting. The difference is consistent with spatial dependence making random splits easier, but also reflects the different geographic transfer task imposed by grouped holdouts. It does not by itself quantify every source of leakage. Unless explicitly identified as random-split or province-held-out results, the reported validation uses DSD grouping; embedding-stage separation remains an unresolved qualification.

\subsection{Geographic Generalisability Varies Across Provinces}

\begin{figure}[t!]
    \centering
    \includegraphics[width=\columnwidth]{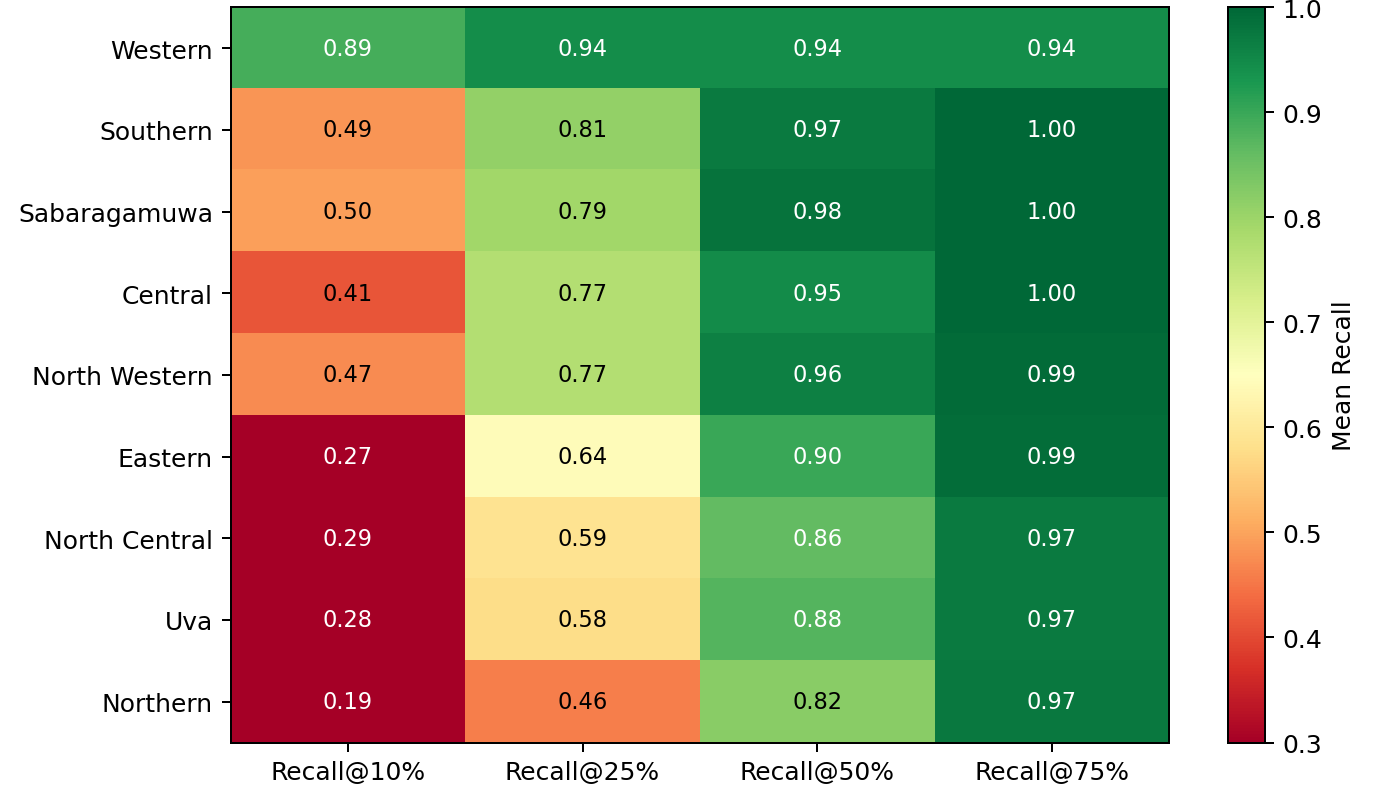}
    \caption{Province-level generalisation under leave-one-province-out evaluation
    (Random Forest). Performance varies considerably across Sri Lanka's nine
    provinces.}
    \label{heatmap}
\end{figure}

Figure~\ref{heatmap} reports province leave-one-out results for the Random Forest
model. Performance varies considerably and reveals structurally meaningful patterns.
Western Province achieves the highest recall (Recall@25\% $=0.944$), followed by Southern (0.809) and Sabaragamuwa (0.793). Northern Province performs worst (0.461); North Central (0.588) and Uva (0.578) also underperform. These are within-province ranking results and should not be compared directly with the national NTL baseline without evaluating that baseline on the same province holdouts. Differences in infrastructure, livelihoods, conflict history, and phone coverage are possible explanations, not mechanisms identified by this analysis.

At Recall@75\%, Northern Province reaches 0.97. A larger selection set is easier to recover, so this improvement should be read against the random-selection reference of 0.75. High recall at a broad budget does not remove the poor-tail generalisation failure.

\subsection{Targeting Performance Across Data Modalities}

\begin{table*}[t!]
    \centering
    \caption{Mean Recall@$k$ = Precision@$k$ (percent) against PC1 at fixed-count targeting thresholds, as reported across 1,000 DSD-grouped repetitions. ``Combined'' includes CDR, tabular RS, and CNN embeddings. Values are descriptive comparisons; paired split-level uncertainty and an embedding-only ablation are not available.}

    \label{tab:overlap}
    \begin{tabular}{llrrrr}
    \toprule
    Data type & Model & Poorest 25 DSDs & Poorest 50 DSDs &
    Poorest 100 GNDs & Poorest 500 GNDs \\
    \midrule
    \multicolumn{6}{l}{\textit{Combined: CDR + RS + CNN}} \\
    & Random Forest  & 86\% & 91\% & 55\% & 65\% \\
    & XGBoost        & 86\% & 92\% & 54\% & 64\% \\
    & Decision Tree  & 84\% & 90\% & 43\% & 60\% \\
    \midrule
    \multicolumn{6}{l}{\textit{RS only}} \\
    & Random Forest  & 69\% & 77\% & 44\% & 51\% \\
    & XGBoost        & 69\% & 72\% & 43\% & 49\% \\
    & Decision Tree  & 66\% & 75\% & 42\% & 60\% \\
    \midrule
    \multicolumn{6}{l}{\textit{CDR only}} \\
    & Random Forest  & 67\% & 79\% & 34\% & 49\% \\
    & XGBoost        & 72\% & 81\% & 38\% & 53\% \\
    & Decision Tree  & 68\% & 80\% & 29\% & 48\% \\
    \bottomrule
    \end{tabular}
\end{table*}

Table~\ref{tab:overlap} explicitly reports both targeting metrics for the modality comparison: each percentage is simultaneously mean recall and mean precision because the sets have equal size. For the 25 poorest DSDs by PC1, Random Forest's reported overlap is 86\% for combined inputs, 69\% for RS-only, and 67\% for CDR-only. These are model--PC1 results, not model--HCI results. The combined configuration also contains CNN embeddings, so its advantage cannot be attributed solely to adding CDR.

For the poorest 100 GNDs, the corresponding Random Forest values are 55\%, 44\%, and 34\%; the combined--RS difference is 11 percentage points, compared with 17 points for the 25-DSD threshold. Because these cut-offs select different fractions of units, they do not establish a scale-independent marginal benefit. Comparisons need identical evaluation units, DSD aggregation rules, and paired splits to support inferential claims. The reported summaries do not provide uncertainty for these differences or establish modality-specific superiority within urban and rural subgroups.

\subsection{Spatial Distribution of Predicted Poverty}

Figure~\ref{map} compares the spatial distribution of the Random Forest predicted
poverty score against the PC1 reference at GN division level. Both maps use the
same continuous percentile scale (0--100). The predicted and reference
distributions show strong visual correspondence: the northern and north-central
interior regions are flagged as high-poverty by both the model and PC1, consistent
with the historically deprived dry zone. The Western Province coastal corridor and
southern lowlands appear wealthy in both maps, reflecting the relative prosperity of
Sri Lanka's most urbanised and commercially active regions.

The most visible divergence occurs in the Northern Province, where the model assigns
moderate-to-high poverty scores but with less spatial precision than the reference map.
This visual pattern is compatible with the province holdout results, but neither map correspondence nor geographic coherence rules out statistical artefacts or validates household targeting.

\subsection{Model Performance Across Urban and Rural Contexts}

The reported within-subset ranking correlations across 200 DSD-grouped repetitions are higher in rural areas ($\rho=0.756$) than urban areas ($\rho=0.516$). All 1,925 GNDs below the PC1 training threshold of $-3$ are rural. This label distribution and a restricted range of urban PC1 values may contribute to the difference, but the comparison does not identify its cause. In particular, absence of urban positive labels under this proxy does not imply absence of urban poverty. Urban targeting requires a reference measure that captures deprivation within cities.

\subsection{Spatial Isolation Is Associated with Prediction Error}
\label{sec:isolation}

\begin{figure}[t!]
    \centering
    \includegraphics[width=\columnwidth]{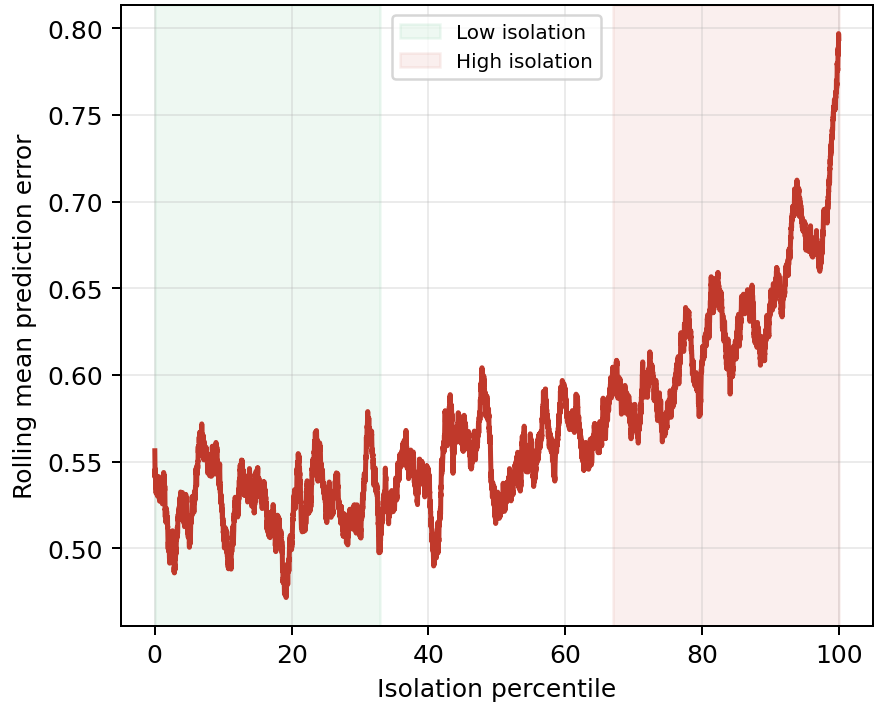}
    \caption{Rolling mean prediction error across the spatial isolation spectrum.
    Each GN division is ranked by its isolation index $I_i$, the absolute difference
    between its PC1 value and the mean PC1 of its eight nearest geographic
    neighbours. Error fluctuates around 0.50--0.55 at low isolation and reaches approximately 0.80 at the upper endpoint. Green and red shading identify the lowest and highest terciles; the rolling curve is not monotonic.}
    \label{isolation}
\end{figure}

To examine errors in socioeconomically atypical units, including poor enclaves surrounded by wealthier neighbours and the reverse, we define a spatial isolation index:

\begin{equation}
I_i = \left| \text{PC1}_i - \frac{1}{K}\sum_{j \in \mathcal{N}(i)} \text{PC1}_j \right|
\end{equation}

where $\mathcal{N}(i)$ denotes the $K = 8$ nearest geographic neighbours of GN
division $i$ identified by haversine distance on centroid coordinates. Higher values
of $I_i$ indicate that a GN division's socioeconomic status diverges sharply from
its immediate surroundings.

The highest-isolation tercile has a reported mean absolute prediction error 10.8\% above the lowest tercile. The overall rank association is small (Spearman $\rho=0.079$, $n=13{,}985$); its size is more informative than a small nominal $p$-value when observations are spatially dependent. Figure~\ref{isolation} shows local increases in error, particularly toward the upper end of the isolation distribution. Because $I_i$ uses PC1, it is a retrospective diagnostic requiring reference labels, not an observable deployment-time warning score. It also includes both poor enclaves and unusually wealthy units, so the aggregate result alone does not measure exclusion of poor enclaves.

The modality comparison reports isolation--error correlations of 0.178 for RS-only, 0.134 for CDR-only, and 0.167 for combined inputs. The descriptive RS--CDR difference is 0.044. These modality-specific summaries should not be pooled with the overall $0.079$ estimate without reconciling their prediction aggregation and error definitions. No paired interval or test for the difference is available. A suitable comparison would resample DSDs jointly across modalities using the same held-out predictions, accounting for residual dependence across boundaries where necessary.

A stronger RS-only association is consistent with spatially smooth landscape features obscuring local socioeconomic contrasts. It does not establish RS smoothness as the dominant cause or exclude CDR tower aggregation, proxy error, differences in coverage, or model misspecification. Both the isolation index and prediction error depend on PC1, creating an additional potential source of association. Label-independent diagnostics and controlled feature-resolution comparisons are needed before extending this mechanism claim to other countries or pipelines.

To assess whether the bias could be reduced through feature engineering, we augmented
the RS feature set with neighbourhood delta features, explicit representations of how
each GND's RS values diverge from its geographic neighbours:

\begin{equation}
\Delta f_i = f_i - \frac{1}{K}\sum_{j \in \mathcal{N}(i)} f_j
\end{equation}

for each RS feature $f$. Adding these delta features produced negligible improvement
on held-out data: isolation-error correlation changed by $\Delta\rho = +0.001$,
high-isolation error inflation changed by $+0.1$ percentage points, and Recall@25\%
improved by only $+0.17$ percentage points. This negative result applies to the tested features and model; it neither demonstrates that the ensemble already captures local contrast nor establishes that an architectural change is necessary. Candidate mitigations and the evidence needed to evaluate them are discussed in Section~\ref{sec:discussion}.

\section{Discussion}
\label{sec:discussion}

The models recover much of the PC1 poor tail, but performance depends on where they are tested. Random splits give higher scores than DSD holdouts, and the province results expose weaknesses that a national average would hide. These differences matter when a map is used to decide which communities receive attention.

\subsection{What the Targeting Results Establish}
The main limitation is the reference measure. A model can reproduce PC1 accurately and still miss communities experiencing consumption poverty. PC1 recovers only 17 of the 25 DSDs selected by HCI. This overlap and the model's Recall@25\% of 0.83 concern different units and cut-offs; they cannot be combined into an estimate of model--HCI accuracy. Direct comparison on matched DSDs, accounting for HCI uncertainty where available, is needed to assess that accuracy.

Our metrics also give every administrative unit equal weight. An allocation programme would need to account for population, intervention costs, and eligibility. If eligibility is fixed while the budget varies, precision and recall would measure different errors, unlike the equal-sized sets used here. The scores predict an asset-deprivation label; they do not identify causes of poverty or estimate how much an intervention would help.

\subsection{Spatial Validation and Remaining Tests}
The 4.1-percentage-point difference between random and grouped splits shows how validation design changes the apparent performance. DSD grouping reduces overlap between nearby training and test units, but dependence can persist across boundaries. The CNN training procedure must also be checked before representation-stage leakage can be ruled out.

We have not tested whether the available training data are sufficient. Spatial learning curves would address this by holding out the same DSDs and increasing the number of training DSDs, with preprocessing fitted within each training subset. Recall and subgroup errors across repeated outer splits would show where additional observations help. A plateau would apply to the tested model and sampling pattern. Sensitivity to the PC1 training threshold, comparisons with stronger image models, and a matched ablation of CNN embeddings also remain open.

\subsection{Isolation Errors and Change Over Time}
Neighbourhood delta features barely changed the isolation results. This gives little guidance on which remedy would work. Higher-resolution inputs or additional labels from atypical communities are possible next steps. Multiscale or spatial-attention models deserve comparison, but could also smooth over the contrasts we want to preserve. Any improvement should be tested on the same spatial holdouts, separating poor enclaves from wealthy outliers and checking against a reference that does not depend on PC1 where possible.

The data describe 2012--2013. Since then, changes in phone use, network coverage, settlement, prices, and livelihoods may have altered the relationships the models learned. Updating the map would require recent survey observations, compatible boundaries, and temporal holdouts. Retraining or transfer learning could then be evaluated against those observations. We have not tested either approach, and these results do not establish present-day targeting performance.

\subsection{Ethics and Data Governance}
\label{sec:ethics}
CDRs reveal sensitive patterns of communication and movement. Removing identifiers does not prevent all re-identification \cite{demontjoye2013unique}, and tower or GND aggregation provides no formal privacy guarantee. Access granted by an operator also does not establish subscriber consent for poverty inference. The original consent or waiver, ethics-review determination, access controls, and retention arrangements are not documented here and need to be established to assess this study's data governance.

For future use, agreements should limit the purpose of analysis, restrict access, set aggregation and retention rules, and provide independent oversight. They should prohibit surveillance and commercial profiling and specify the reviewed basis for secondary use. Differential privacy is one option, but would require a defined privacy unit, contribution limits, privacy budget, and accounting for repeated releases. Its effect on targeting small communities would need testing. Neither differential privacy nor the proposed alternatives of federated computation and synthetic releases were evaluated in this study; the latter two do not by themselves guarantee privacy.

Representation is a separate concern. People without phones, shared-phone users, and subscribers to other operators may be poorly represented. Aggregating their neighbours' records does not correct that omission. Ethnicity and demographic features need a clear justification and an audit for discriminatory effects. Area scores should not be treated as household poverty labels. RS-only models avoid dependence on proprietary phone records, but still require checks for unequal errors and the risks of publishing sensitive community profiles.

\subsection{Use in Allocation and Reproducibility}
In a possible Sri Lankan application, statistical agencies, DSD administrators, programme staff, and community representatives would first agree on eligibility, costs, and the allocation budget. After validation with current data, rankings could guide survey visits or outreach. Recent household evidence and community consultation would then inform decisions, with a route for correcting omissions and appealing them. This process has not been evaluated here. A low model score should not alone exclude a household from support, and outreach should include areas with weak coverage or known model failures. Monitoring should track exclusion, geographic disparities, and programme outcomes, as well as potential stigma or misuse of the maps.

Replication depends on access to CDRs. An operator-managed analysis environment with reviewed aggregate exports could support replication without releasing raw records. Feature definitions, preprocessing code, split identifiers, hyperparameters, and approved aggregate outputs would also help others assess the work. These are possible arrangements rather than confirmed safeguards or release commitments. Where CDR access is unavailable, the RS-only results provide a useful comparison, with lower PC1 overlap but fewer access barriers.

\bibliography{aaai25}

\end{document}